\documentclass{JHEP}
\preprint{
CERN-TH/2001-093\\
DAMTP-2001-20\\
T/01-032\\
hep-th/0104023 \\
}
\title
{Cosmological Evolution on Self-Tuned Branes and the Cosmological Constant}
\author
{ Ph. Brax \\
Theoretical Physics Division, CERN\\
CH-1211 Geneva 23.
\thanks{ On leave of absence from  Service de Physique Th\'eorique, 
CEA-Saclay F-91191 Gif/Yvette Cedex, France.}\\
\email{philippe.brax@cern.ch}}
\author
{ A. C. Davis\\
DAMTP, Centre for Mathematical Sciences \\ Cambridge University,
Wilberforce Road, Cambridge, CB3 0WA, UK.\\ 
\email{a.c.davis@damtp.cam.ac.uk}}

\abstract{
We consider the cosmological evolution of a bulk scalar field and ordinary 
matter living on the brane world in the light of the constraints imposed by
the  matter dominated cosmological evolution and a small cosmological constant now. We  rule out models with a self-tuned minimum of the four dimensional potential as they would lead to rapid oscillations of the Hubble parameter now.
A more natural framework is provided by supergravity in singular spaces where
the brane coupling and the bulk potential are 
related by supersymmetry, leading  to a four dimensional run-away potential.
For late
times we obtain an accelerating universe due to the breaking of
supersymmetry on the brane with an acceleration parameter
of $q_0=-4/7$ and associated equation of state $\omega=-5/7$. 
}
\begin{document}

\section{Introduction}

Recent experimental results seem to indicate that the universe undergoes
a phase of late inflation \cite{supernova}. Such an accelerating universe can 
be described from the four dimensional point of view using quintessence 
models \cite{quintessence}. However, one of the 
drawbacks of this approach is that the cosmological constant problem is 
assumed to be solved by an as yet unknown mechanism.
Lately it has been advocated that a vanishing cosmological constant can be 
easily obtained using five dimensional brane-world models\cite{dimopoulos}. 
In that context the vacuum energy
on the brane curves the fifth dimension while preserving a flat brane-world. 
In particular this situation is typical in singular space supergravity where 
flat brane-worlds are a consequence of the BPS equations\cite{kallosh}. 

It is a challenge to produce an accelerated universe in brane-world scenarios.
This has been recently achieved by breaking supersymmetry on the brane-world 
\cite{us}.
The cosmological solutions obtained along these lines correspond to an 
accelerating universe with a deceleration parameter $q_0=-4/7$, within the 
experimental ballpark, and a corresponding equation of state, $\omega=-5/7$.
It is therefore highly relevant to couple the previous matter-free scenarios 
to matter on the brane-world. 

The five dimensional brane cosmology is commonly described using the
ansatz for the metric
\begin{equation}
ds^2=-N^2(t,x_5)dt^2+ A^2(t,x_5)dx^idx_i + B^2(x_5,t)dx_5^2.
\label{metric1}
\end{equation}
When the three function $N, \ A$ and $B$ are both time and $x_5$-dependent
this leads to the violation of Lorentz invariance in the bulk\cite{ishi}. In 
that 
context it is
unlikely that a four dimensional effective action can be obtained as a 
consistent truncation of the five dimensional  degrees of freedom. In
particular the meaning of the four dimensional Planck mass is not clear if one
cannot obtain a Poincar\'e invariant description of the four dimensional 
physics.

Moreover it is often argued that one should also impose that the radius
of the fifth dimension is stabilized. This is to comply with the five-dimensional
rephrasing of the hierarchy problem in the Randall-Sundrum scenario \cite{RS}.
This amounts to imposing
that $B$ is time independent. From the cosmological point of view this does
not seem to be compulsory as explicit solutions with a time-varying fifth 
dimension exist in singular space supergravity  and  lead to an accelerated 
universe on the brane-world \cite{us}. 
In the following we shall use a particular case of (\ref{metric1}) with two 
special features. 
With this choice  we will be able to recover the cosmological solutions of supergravity in singular
 spaces.  Similarly the four dimensional dynamics 
involving the scale factor $a$ , the matter density $\rho_m$ and the
restriction to the brane-world of the bulk scalar fields $\phi$, is closed
in the case of bulk supergravity.
We do not require that such a four dimensional description of the brane-world 
follows from
a four dimensional effective action for $a$ and $\phi$.
We shall adopt a phenomenological attitude and compare the closed dynamics issued from the
five dimensional model with the usual four dimensional Klein-Gordon, Friedmann and conservation equations.
This will lead to the identification of an effective Newton constant which appears to be
dependent on the scalar field $\phi$, analogous to Brans-Dicke theory.
In the radiation dominated era this dependence disappears
and the effective Newton constant is time independent. In the matter 
dominated era the effective Newton constant is slowly varying in time.   
 
We will examine two contrasting situations. As already mentioned the broken 
supergravity case
yields a phase of late acceleration. From the four dimensional point of view it is a model
with an exponential runaway potential. Another interesting possibility is that
the potential possesses a self-tuned minimum, i.e. for each value of the brane 
tension  there exists a value of the scalar field representating the vacuum 
state.
These two situations are typical and will be investigated in depth. 

The plan of the paper is as follows. In section 2 we set up brane world
cosmology, describing the coupling between ordinary matter on the brane
and the bulk scalar field. The cosmological dynamics is considered in section
3 where we derive the full Friedmann and Klein-Gordon equations, showing the
special limits of the Randall-Sundrum model \cite{RS} and the self-tuned cosmological constant on a 
brane \cite{dimopoulos}, for recent work see 
\cite{grojean&hollowood,grojean&binetruy,selftuning}. This is applied to the 
cosmological evolution of
brane-worlds with a self-tuned vacuum  in section 4. All periods of the evolution of the
universe are considered in this section. In the matter dominated phase, our
solution gives rise to an oscillating universe. Requiring that the scalar field energy
on the brane-world becomes of the same order of magnitude as the matter energy
in the recent past, i.e. in order to accommodate a non-negligible dark energy component,
we find  that the four dimensional mass of the scalar field has to be tuned. In turn
this leads to very rapid oscillations of the Hubble parameter now, ruling out
the model.  In section 5
we consider supergravity in singular spaces \cite{kallosh}, with no matter on
the brane. We consider the case of supersymmetry breaking on the brane and 
recover the accelerating universe of \cite{us}. The cosmological evolution
in the presence of brane world matter in considered in section 6. In the
matter dominated era the Newton constant is time dependent, though it is 
strictly constant in the radiation era. Since other
couplings are likely to evolve with time in a similar manner, this
may well evade detection. 
Finally we show that the universe can be in a phase of late acceleration 
where
the scalar field energy dominates now. Imposing that the scalar field energy is
subdominant in the radiation and matter era we find that the amount of 
supersymmetry breaking on the brane has to be 
fine-tuned and corresponds to the critical energy density now. 
This leads to a reformulation of the cosmological constant problem in 
brane-world scenarios.

\section{Brane Cosmology}

We consider our universe to be one of the boundaries of  five dimensional space-time.
The embedding is chosen such that our brane-world sits at the origin of the fifth dimension.
We impose a $Z_2$ symmetry along the fifth dimension and identify $x_5$ with $-x_5$. 
Our brane-world carries two types of matter, the standard model fields at sufficiently large energy
and ordinary matter and radiation at lower energy. We also assume that gravity propagates
in the bulk where a scalar field $\phi$ lives. This scalar field couples to the standard model fields
living on the brane-world. At low energy when the standard model fields have condensed and
the electro-weak and hadronic phase transitions have taken place, the coupling of the scalar field
to the brane-world realizes the mechanism proposed in \cite{dimopoulos} with a self-tuning
of the brane tension.  
In this section we derive the brane cosmology equations describing the
coupling between ordinary matter on the brane and a scalar field in the bulk.
Our derivation is similar to \cite{carsten}.

Consider the bulk action
\begin{equation}
S_{bulk}=\frac{1}{2\kappa_5^2}\int d^5 x \sqrt {-g_5}(R-\frac{3}{4}((\partial\phi)^2+U(\phi)))
\end{equation}
where $\kappa_5^2=1/M_5^3$ and the boundary action
\begin{equation}
S_{B}=-\frac{3}{2\kappa_5^2}\int d^4 x \sqrt{ -g_4}U_B(\phi_0)
\end{equation}
where $\phi_0$ is the boundary value of the scalar field.
The Einstein equations read
\begin{equation}
G_{\mu\nu}\equiv R_{\mu\nu}-\frac{1}{2}R g_{\mu\nu}=T_{\mu\nu} +\delta_{x_5}T^B_{\mu\nu}
\end{equation}
where $T_{\mu\nu}$ is the bulk energy-momentum tensor and $T^B_{\mu\nu}$
is the boundary contribution. 
The bulk term is
\begin{equation}
T_{\mu\nu}=\frac{3}{4}(\partial_{\mu}\phi\partial_{\nu}\phi -\frac{1}{2}g_{\mu\nu}(\partial\phi)^2)
-\frac{3}{8}g_{\mu\nu}U
\end{equation}
and the boundary term
\begin{equation}
T^B_{ab}=-\frac{3}{2} g_{ab}U_B(\phi)
\end{equation}
with $a,b=0\dots 3$.
Following the self-tuning proposal we interpret $U_B$ as arising from a direct coupling
$U_B^0$ to the brane degrees of freedom, i.e. the standard model fields $\Phi^i$. The vacuum energy generated
by the $\Phi^i$'s yields the effective coupling
\begin{equation}
\frac{3U_B}{2\kappa_5^2}=<V(\Phi)>U_B^0
\end{equation}
where the dimension four potential  $V(\Phi)$ represents all the contributions due  the fields $\Phi^i$ after
inclusion of condensations, phase transitions and radiative corrections.

We also consider that ordinary matter lives on the brane with a diagonal energy momentum tensor
\begin{equation}
T_{b}^{a\ matter}=\hbox{diag}(-\rho_m,p_m,p_m,p_m)
\end{equation}
and an equation of state $p_m=\omega_m \rho_m$.
We consider the metric
\begin{equation}
ds^2=a^2(t,x_5)b^2(x_5)(-dt^2+dx_5^2)+ a^2(t,x_5)dx_idx^i.
\label{metric}
\end{equation}
This is a particular subset of (\ref{metric1}) motivated by the possibility
of  retrieving  conformally flat metrics when matter is not present on the brane.  
Notice that $\dot b=0$ and the only time dependence in the metric appears in $a$.

Einstein's equations are
\begin{equation}
\frac{\partial_{t}^2a}{a}-2\frac{(\partial_{x_5} a)^2}{a^2}-\frac{\partial_{x_5} a\partial_{x_5} b}
{ab}=-\frac{a^2b^2}{3}T^5_5
\label{55}
\end{equation}
and
\begin{equation}
-\frac{\partial_t\partial_{x_5}a}{a}+2\frac{\partial_{x_5} a\partial_{x_5} a}{a^2}+
\frac{\partial_{t} a\partial_{x_5} b}{ab}=-\frac{a^2b^2}{3}T^{0}_5
\label{05}
\end{equation}
along the normal direction. The other components of Einstein's equations lead to the 
jump conditions on the brane-world.

In the following we will only be interested in the dynamics of the brane-world.
This is achieved by restricting the equations of motions to the brane-world. The particular
choice (\ref{metric}) for the metric allows us to write down three 
independent equations
for $\phi$, $\rho_m$ and $a$ on the brane-world. These equations are the analogues of the usual
four dimensional Friedmann, Klein-Gordon and conservation equations.
It is convenient to use proper time and distances on the brane.
The proper time is defined by
\begin{equation}
d\tau =ab \vert_{0}dt 
\end{equation}
and the normal vector to the brane
\begin{equation}
\partial_n=\frac{1}{ab}\vert_{0}\partial_{x_5}.
\end{equation}
We will denote the normal derivative by prime and by dot the proper time derivative.
From now on all the identities will be explicitly taken to be on the brane-world at $x_5=0$.

The Israel conditions on the extrinsic curvature lead to the following 
boundary conditions  
\begin{equation}
\frac{a'}{a}=-\frac{1}{6}\rho ,\ \frac{b'}{b}=\frac{1}{2}(\rho+p)
\label{a}
\end{equation}
where $\rho$ and $p$ are the total energy density and pressure on the brane
\begin{equation}
\rho=\rho_m+\frac{3}{2}U_B,\ p=p_m -\frac{3}{2}U_B.
\end{equation}
The combination $\rho +p$ does not involve the scalar field $\phi$. 
In the absence of matter $b$ can be chosen to be constant in the brane vicinity.
Similarly the $\phi$ boundary  condition reads
\begin{equation}
\phi'=\frac{\partial U_B}{\partial \phi}.
\label{phi}
\end{equation}
These boundary conditions have been obtained using the $Z_2$ symmetry in an explicit way.
They will allow us to eliminate  spatial derivatives from the dynamical 
equations.

The derivation of the conservation equation requires 
\begin{equation}
r\equiv - T^0_5=\frac{3}{4}\phi'\dot \phi.
\end{equation}
Using the boundary conditions we get
\begin{equation}
r=\frac{3}{4}\dot U_B.
\end{equation}
Restricting (\ref{05}) to the brane and using the boundary conditions (\ref{a}) we get
the conservation equation
\begin{equation}
\dot \rho =-3 H (p+\rho) +2r
\end{equation}
where
\begin{equation}
H=\frac{\dot a}{a}
\end{equation}
is the Hubble parameter of the brane-world.
The last term comes from the coupling with the bulk and involves the time derivative of the scalar field.
It cancels the time variation of the scalar energy density $3U_B(\phi)/2$ on the brane.
Finally we obtain 
\begin{equation}
\dot \rho_m =-3 H (\rho_m+p_m)
\end{equation}
as in four dimensional cosmology implying that
\begin{equation}
\rho_m=a^{-3(1+\omega_m)}.
\end{equation}
Notice that matter does not leak out of the
brane-world. 

The Friedmann equation is obtained from  (\ref{55}). 
The  component of the  energy momentum tensor that is needed    
\begin{equation}
q\equiv T^5_5= \frac{3}{8}(\dot\phi^2+\phi'^2- U)
\end{equation}
involves the pressure due to the scalar field in the bulk.
Using (\ref{phi}) this reads 
\begin{equation}
q=  \frac{3}{8}(\dot\phi^2+\nabla U_B^2 - U).
\end{equation}

Similarly from (\ref{55}) and (\ref{a}) we get
\begin{equation}
\frac{\ddot a}{a}+(\frac{\dot a}{a})^2=-\frac{1}{36}\rho (\rho+3p) -\frac{q}{3}.
\label{fr}
\end{equation}
The Friedmann equation is a first integral obtained by putting $a=e^{c}$.
Substituting in (\ref{fr}) we obtain 
\begin{equation}
\frac{d(e^{4c}H^2)}{dc}=-\frac{2e^{4c}}{3}(\frac{\rho (\rho+3p)}{12}+q).
\end{equation}
Using the conservation equation this can be written as
\begin{equation}
\frac{d(e^{4c}H^2)}{dc}=\frac{1}{36}\frac{d(\rho^2 e^{4c})}{dc}- \frac{e^{4c}}{9}\frac{r \rho}{H}-\frac{2q}{3}
\end{equation}
which can be integrated to yield the Friedmann equation
\begin{equation}
H^2=\frac{\rho^2}{36}-\frac{2}{3}Q-\frac{1}{9}E + \frac{\cal A}{a^4}
\end{equation}
where ${\cal A}$ is a constant.
The functions  $Q$ and $E$ satisfy the differential equations
\begin{equation}
\dot Q +4 HQ= Hq
\end{equation}
and
\begin{equation}
\dot E +4 HE= \rho r.
\end{equation} 
Notice that the non-conventional $\rho^2$ term is retrieved. The functions $Q$ and $E$ capture
the whole dynamics induced by the coupling of the bulk scalar field to the 
brane \cite{carsten}.

\section{Cosmological Dynamics}

In this section we will give the general form of the Friedmann and Klein-Gordon equations
when the scalar field in the bulk couples to the brane-world. In particular this will enable us to 
discuss some of the non-conventional effects of the brane-world cosmology. More details
will be given in the ensuing sections where we apply the formalism to the case of a brane with a self-tuned vacuum
and to supergravity in singular spaces.

\subsection{No Matter on the Brane}

Let us first study the matter-less situation $\rho_m=0$. 
We need to solve the differential equations for $Q$ and $E$. 
The differential equation for E involves
\begin{equation}
\rho r= \frac{9}{16} \frac{dU_B^2}{d\tau}
\end{equation}
implying that
\begin{equation}
-\frac{E}{9}=-\frac{1}{16a^4}\int d\tau a^4 \frac{dU_B^2}{d\tau}.
\end{equation}
Similarly we find that
\begin{equation}
-\frac{2}{3}Q=-\frac{1}{16}(\dot\phi^2+(\nabla U_B)^2-U) +\frac{1}{16a^4}\int d\tau a^4 \frac{d}{d\tau }.
(\dot \phi^2+ (\nabla U_B)^2-U)
\end{equation}
This leads to 
\begin{equation}
H^2=-\frac{1}{16 a^4}\int d\tau \frac{da^4}{d\tau}(\dot \phi^2-2V)+ \frac{\cal A}{a^4}
\label{H}
\end{equation}
where we have identified  the effective four dimensional potential
\begin{equation}
V=\frac{U_B^2-(\nabla U_B)^2+U}{2}
\end{equation}
combining the bulk potential and the brane coupling.
The dynamics is completely specified once the Klein-Gordon is written, 
\cite{wands}
\begin{equation}
\ddot \phi + 4H\dot \phi =-\nabla V +\Delta\Phi_2
\end{equation}
where 
\begin{equation}
\Delta \Phi_2\equiv \phi''-\nabla^2 U_B \nabla U_B
\end{equation}
is the contribution due to the five-dimensional dynamics of the scalar field and measures the
scalar energy momentum loss from the brane to the bulk.
This contribution cannot be derived from the four-dimensional point of view only.
It necessitates a global analysis of the equations of motion in five dimensions.

In the following we shall encounter two typical situations. First of all in the 
cases where a constant scalar field is a solution on the brane, as in the 
Randall-Sundrum and 
self-tuned brane scenarios, the loss parameter $\Delta \Phi_2$ vanishes. 
Secondly in the matter-less supergravity and broken supergravity cases the loss
parameter vanishes altogether too. This is due to the global validity of the boundary condition (\ref{phi})
throughout the bulk.
In the following we will assume that the loss parameter vanishes
\begin{equation}
\Delta\Phi_2=0.
\end{equation} 
Physically this may be interpreted by considering that the brane-world is not an infinitely thin
boundary wall but possesses an infinitesimal extension in the $x_5$ direction over which
the boundary condition (\ref{phi}) should be valid. In that case one can take derivatives of
(\ref{phi}) in the $x_5$ direction and obtain $\Delta\Phi_2 =0$. Notice that this
sets the time evolution of $\phi$ on the brane-world. It is by no means obvious that
a global solution with this boundary evolution exists throughout the bulk. It is the case in
supergravity and we shall assume that this also stands more generally.

The equation (\ref{H}) differs greatly from the four dimensional Friedmann equation 
\begin{equation}
H^2=\frac{8\pi G_N}{3}(\frac{\dot\phi^2}{2}+V)
\end{equation}
as it is an integral equation. In the slow-roll approximation where the time variation
of $\phi$ is negligible compared to the time variation of $a$ one can rewrite (\ref{H})
as 
\begin{equation}
H^2=-\frac{1}{8}(\frac{\dot \phi^2}{2}-V)
\end{equation}
where we have assumed that the dark radiation term vanishes.
It is immediately apparent that the brane-world dynamics does not mimick the
four dimensional case as it is the pressure $\dot \phi^2/2-V$ which appears
in the brane-world case. This springs from the fact that $T^5_5$ is the bulk pressure
and not the bulk energy density.

\subsection{Including Matter}

The effect of including matter on the brane is twofold.
First there is a direct contribution coming from $\rho^2/{36}$.
Then  there is a new contribution to $E$ from $r \rho_m$. 
This leads to the complete Friedmann equation
\begin{equation}
H^2= \frac{\rho_m^2}{36}+ \frac{1}{12}U_B\rho_m-\frac{1}{16 a^4}\int d\tau \frac{da^4}{d\tau}(\dot \phi^2-2V)-\frac{1}{12 a^4}\int d\tau a^4 \rho_m \frac{dU_B}{d\tau}.
\label{fried}
\end{equation}
The first term is responsible for the non-conventional cosmology in the early universe. The linear term in $\rho_m$ will lead to the matter and radiation dominated eras. It involves 
an effective Newton constant
\begin{equation}
\frac{8\pi G_N}{3}\equiv \frac{\kappa_5^2 U_B}{12}
\end{equation}
where we have reintroduced the dimensionful parameter $\kappa_5^2$.
The dynamical contribution of the scalar field
is encapsulated in the two remaining integrals. The first integral involves the pressure of the scalar field.
The last integral is only relevant when the effective Newton constant is time dependent.

In the same fashion  the 
Klein-Gordon equation is modified due to the boundary conditions (\ref{a})
\begin{equation}
\ddot \phi + 4H\dot \phi +\frac{1}{2}(\frac{1}{3}-\omega_m)\rho_m\nabla U_B=-\nabla V.
\label{KG}
\end{equation}
Notice that the new contribution vanishes for the radiation fluid, and we have
assumed the loss parameter vanishes, as discussed previously.

Let us comment on some particular cases.
First of all when
\begin{equation}
U_B=\lambda,\ U=-\Lambda
\end{equation}
together with
\begin{equation}
\Lambda=\lambda^2
\end{equation}
one retrieves the Randall-Sundrum case \cite{RS} with no scalar field in the 
bulk and a vanishing four-dimensional
potential
\begin{equation}
V_{RS}=0.
\end{equation}
The effective Newton constant is
\begin{equation}
\frac{8\pi G_N}{3}=\frac{\kappa_5^2 \lambda}{12}
\end{equation}
where we have reintroduced the five dimensional scale. The tuned
Randall-Sundrum scenario has been used as a paradigm for theories with
a non-compact extra dimension; it has been demonstrated to reproduce
the usual gravitational interactions, with any deviations being undetectable
to present levels of accuracy. Indeed the Randall-Sundrum model has initiated
a wealth of studies on the cosmology of extra dimensional theories. In 
\cite{bine} it was shown that there are modifications to the Friedmann
equation with extra expansion at very early times, as in (\ref{fried}). 
However, the RS 
model relies on the bulk cosmological constant and the brane tension
being tuned as above. Hence it does not solve the long-standing problem of 
the cosmological constant. However, see \cite{cosconst} for various 
suggestions.

Similarly the self-tuned brane scenario of \cite {dimopoulos} corresponds to
\begin{equation}
U_B=Te^{\pm \phi}, \ U=0
\end{equation}
where $T$ comes from $<V(\Phi)>$. This model 
leads to an identically vanishing four-dimensional potential
\begin{equation}
V_{ST}=0.
\end{equation}
Despite the presence of a varying tension the effective potential vanishes altogether.
It is well-known that this model has been proposed as a paradigm for a solution
of the cosmological constant problem. Indeed whatever the tension of the brane, i.e.
the vacuum energy of our brane-world, one find flat solutions
\begin{eqnarray}
\phi&=&\phi_0 \pm \ln( 1-\frac{\vert y\vert}{\vert y_c\vert})\nonumber \\
a&=& (1-\frac{\vert y\vert}{\vert y_c\vert})^{1/4}\nonumber \\
\end{eqnarray}
where $dy=adx_5$.
The fact that the brane-world is flat leads to a vanishing  four dimensional
cosmological constant. As such, the cosmological constant problem is solved
in these models, but they do not explain the observations of a small
cosmological constant. They also lead to singularities in the bulk, see for 
example \cite{grojean&hollowood,grojean&binetruy,sing}.
We will comment on the coupling of this model to matter in the following.

\section{Self-tuned Brane Cosmology }

\subsection{ The Early Universe}
In the most general case $U_B$ and the bulk potential, $U$, are not 
functionally related. 
We will first describe some of the features of this more general case.
Generically the Friedmann equation in (\ref{fried}) results in there being 
four different eras. In the very early universe 
the density satisfies 
\begin{equation}
\rho_m>>\frac{2 U_B}{\kappa_5^2}.
\end{equation}
This means that the matter energy density dominates over the brane tension
due to the scalar field, and thus the first term in (\ref{fried}) is the
most relevant.

In that extreme case (\ref{fried}) gives 
\begin{equation}
\rho_m=\rho_0a^{-3(1+\omega_m)}
\end{equation}
and
\begin{equation}
a(\tau)= \tau^{ 1/3(1+\omega_m)}.
\end{equation}
As $\rho_m$ decreases the linear term becomes dominant.
This will be followed by the usual radiation and matter dominated eras.
Finally the presence of a scalar field with a non-vanishing potential
$V$ can lead to a phase of late acceleration
in a way reminiscent of quintessence models.
Let us examine in detail the radiation and matter dominated eras.
\subsection{Radiation dominated era}

As long as one can neglect the potential $V$, the dynamical equations are
satisfied by 
\begin{equation}
\phi=\phi_0.
\end{equation}
From (\ref{fried}) this is a valid approximation as long as
\begin{equation}
\frac{V(\phi_0)}{U_B(\phi_0)}<<\frac{2}{3}\kappa_5^2\rho_e,
\end{equation}
where $\rho_e$ is the density at equality.
For natural values of $\kappa_5^2\approx 10^{-9}$ GeV$^{-3}$
the ratio of the brane potential to the brane coupling must be
$10^{-48}$ GeV. This is an extreme fine-tuning. Note that, if we had assumed
a larger five-dimensional Planck mass, then an even more extreme fine-tuning
would be required.

In the following we will examine three possible situations. In the first one
the potential on the brane-world vanishes altogether. We have already seen 
that this is
the case of the RS scenario \cite{RS} and of the self-tuned brane model 
\cite{dimopoulos}. 
This is also what happens in five dimensional supergravity in singular spaces, 
where the coupling to the
brane is determined by the bulk superpotential. The second case we will be 
interested in is that of broken supergravity,
where the supersymmetry is broken on the brane-world. This will lead to 
exponentially decreasing potentials in a way reminiscent of quintessence 
models. We discuss this case in detail in the next section.
Another natural possibility, to which we devote the rest of this section, is 
to assume that $\phi_0$ tunes itself to be at the bottom of the potential
\begin{equation}
V(\phi_0)=0.
\end{equation}
This leads to an exact solution of our dynamical equations provided that
\begin{equation}
\nabla V(\phi_0)=0.
\end{equation}
Variations of the potential are assumed to lead to an adjustable minimum. 
In the matter dominated era this equilibrium is disturbed leading to 
oscillations around the minimum, which we will discuss.

\subsection{Matter Dominated Era}
If we perturb $\phi$ around the minimum of the potential such that,
\begin{equation}
\phi=\phi_0+\delta\phi
\end{equation}
to leading order the perturbed dynamics around the minimum of the
potential is described by the Friedmann equation
\begin{equation}
H^2=\frac{U_B(\phi)\rho_m}{12}-\frac{1}{12a^4}\int d\tau  a^4\rho_m\frac{dU_B(\phi)}{d\tau}
\end{equation}
where the integrated terms involving the kinetic energy and the potential are neglected because they are second order in $\delta\phi$.  
The Klein-Gordon equation possesses a crucial destabilizing term
\begin{equation}
\ddot \phi + 4H\dot \phi +\frac{1}{6}\rho_m\nabla U_B=-\nabla V.
\end{equation}
The term proportional to $\rho_m$ implies that $\phi$ cannot remain at the 
bottom of the potential.
In the following we assume for simplicity that the variation of $U_B$ with $\phi$  can be neglected.
It can be easily checked that this assumption is satisfied by the following 
solution.

The field $\delta\phi$ acquires an initial velocity and acceleration according to 
\begin{equation}
\ddot \phi\vert_{\tau_e}+4 H_e\delta\dot\phi\vert_{\tau_e}=-2\alpha {H_e}^2
\end{equation}
where
\begin{equation}
\alpha=\nabla\ln U_B\vert_{\phi_0}
\end{equation}
and $H_e$ is the Hubble constant at equal matter and radiation.
The motion comprises two components, there are free oscillations due to the 
mass
\begin{equation}
m^2(\phi_0)={\nabla^2 V(\phi_0)}
\end{equation}
and a forced motion due to the $\rho_m$ term.
The solution  reads 
\begin{equation}
\delta\phi=2\alpha \frac{H_e^2}{m^2(\phi_0)}(\cos(m(\phi_0)(\tau-\tau_e))-(\frac{\tau_e}{\tau})^2)-
\alpha\frac{H_e^3}{m(\phi_0)^3}\sin(m(\phi_0)(\tau-\tau_e))
\end{equation}
where we have neglected the damping as we  assume that the mass 
$m(\phi_0)$ is much larger than the Hubble parameter at equality.
This leads to 
\begin{equation}
\delta\phi=2\alpha \frac{H_e^2}{m^2(\phi_0)}\cos(m(\phi_0)(\tau-\tau_e))
\end{equation}
soon after equality.
As a result of the oscillations in the perturbation, the potential energy 
oscillates in time 
\begin{equation}
V(\phi)=2\frac{\alpha^2H_e^4}{m^2(\phi_0)}\cos^2(m(\phi_0)(\tau-\tau_e)).
\end{equation}
Similarly, the kinetic energy is also oscillating
\begin{equation}
\dot\phi^2=4\frac{\alpha^2H_e^4}{m^2(\phi_0)}\sin^2(m(\phi_0)(\tau-\tau_e)).
\end{equation}
Now the scalar contribution to the Hubble parameter is
\begin{equation}
-\frac{1}{16 a^4}\int_{\tau_e}^{\tau} d\tau \frac{da^4}{d\tau}(\dot \phi^2 -2V)=
\frac{2\alpha^2 H_e^4}{3m^2(\phi_0)\tau^{8/3}}\int_{\tau_e}^{\tau} d\tau \tau^{5/3} \cos (2m(\phi_0)(\tau-\tau_e))
\end{equation}
whose order of magnitude is bounded by $\alpha^2 H_e^4/4m^2(\phi_0)$. 
At early times in the matter dominated era this term in the Friedmann
equation will be subdominant and we will obtain the usual matter domination.
However, as in quintessence models, there could be a time when the scalar
field dynamics dominate.
Coincidence will happen when
this is of the order of $H^2$, implying  that 
\begin{equation}
\frac{\tau_c}{\tau_e}=O( \frac{2m(\phi_0)}{\alpha H_e}).
\end{equation}
If $m(\phi_0)=10^3$ GeV, this implies that the scalar field will be subdominant
even in the very far future. To obtain coincidence now, and thus an 
accelerating universe, we have to fine-tune $m(\phi_0)=O(10^{-28})$ GeV.
Of course we could always keep $m(\phi_0)$  large, the resulting universe
would remain in the matter dominated era even in the very far future.
As this seems to contradict current experiments we will only consider 
scenarios
where the amount of dark energy due to the scalar field is not negligible now.

Is this scenario realistic? 
Let us examine the oscillations of the fundamental coupling constants such as 
the fine structure constant and Newton's constant.
Due to the oscillation of $\phi$ we find that
\begin{equation}
\frac{\delta G_N}{G_N}=\alpha\delta\phi
\end{equation}
during the matter dominated era. The amplitude of the oscillations is 
\begin{equation}
\frac{\delta G_N}{G_N}=O(\frac{\alpha^2 H^2_e}{m^2(\phi_0)}).
\end{equation}
This is very small. When  $m(\phi_0)=10^{-28}$ GeV and $H_e\sim 10^{-33}$ GeV 
we find that
\begin{equation}
\frac{\delta G_N}{G_N}=O(10^{-10})
\end{equation}
which is of course negligible. 

Similarly
 the coupling of the field
$\phi$ to fields living on the brane springs from 
\begin{equation}
\frac{3}{2\kappa_5^2}\int d^4x\sqrt{-g^{(4)}}U_B(\phi)(1+ \frac{V(\psi)}{<V(\Phi)>})
\end{equation}
where $\psi$ are the fluctuations of the brane-world matter fields.
We find that the variation of the coupling constants $\lambda$ to the standard model fields is
\begin{equation}
\frac{\delta \lambda}{\lambda}=\alpha\delta\phi.
\end{equation}
This is the same as for Newton's constant. 
So we obtain that the oscillations of the scalar fields in the matter dominated era do not lead
to a significant time variation of the fundamental coupling constants.

\subsection{Oscillating Universe}
The late evolution of the universe will
be driven by
\begin{eqnarray}
H^2&=& \frac{U_B(\phi_0)\rho_m}{12}-\frac{1}{16 a^4}\int_{\tau_e}^{\tau} d\tau \frac{da^4}{d\tau}(\dot \phi^2 -2V)\nonumber \\
\ddot \phi&=&-{\nabla V}\nonumber\\
\end{eqnarray}
where the damping term is negligible due to the large value of the mass $m(\phi_0)$. 
The scalar field oscillates freely, as before coincidence
\begin{equation}
\delta \phi= 2\frac{\alpha H^2_e}{m^2(\phi_0)}\cos(m(\phi_0)(\tau-\tau_e)).
\end{equation}
The scale factor will be a solution of the differential equation
\begin{equation}
\ddot y -H_e^2 y (\frac{1}{y^{3/2}}-\frac{4\alpha^2H_e^2}{m^2(\phi_0)}\cos(2m(\phi_0)(\tau-\tau_e))=0
\end{equation}
where $y=a^2$. We retrieve the matter dominated era before coincidence when the oscillatory term is negligible. 
Putting
\begin{equation}
y=(\frac{3H_e \tau}{2})^{4/3}(1+\epsilon(\tau))
\end{equation}
we find that $\epsilon(\tau)$ oscillates in time. 
Indeed it is given by
\begin{equation}
\epsilon(\tau)=\frac{\alpha^2 H_e^4}{m^4(\phi_0)}(\cos(2m(\phi_0)(\tau-\tau_e))-1).
\end{equation}
The resulting Hubble parameter is also oscillating. 
This should have drastic consequences for  our present universe 
where  the matter density is still noticeable. In particular we should observe oscillations of the Hubble parameter
with an inverse    period of $m(\phi_0)=O(10^{-28})$ GeV, which is of the order 400 
seconds. This is excluded experimentally, thus ruling out this class of
models.
In the next section we shall examine a situation where the four-dimensional 
potential does not possess a minimum and therefore does not lead to an 
oscillating universe.

\section{Supergravity in Singular Spaces}
In this section we investigate the case of the recently constructed 
supergravity in singular spaces \cite{kallosh}. This differs from the usual 
five-dimensional supergravity theories since space-time boundaries are taken 
into account. In this case supersymmetry breaking on the brane world 
results in an accelerating universe \cite{us}. We start by reviewing the case 
where there is no matter on the brane before including matter.

\subsection{No Matter on the Brane}

When supergravity in the bulk couples to the boundary in a supersymmetric way 
the Lagrangian is entirely specified by the superpotential
\begin{equation}
U_B=W
\end{equation}
and the bulk potential
\begin{equation}
U=\nabla W^2  -W^2.
\end{equation}
If one considers vector supermultiplets then supersymmetry imposes that
\begin{equation}
W=\xi e^{\alpha \phi}
\end{equation}
where $\alpha=1/\sqrt 3,\ -1/\sqrt 12$, these values arising from the
parametrisation of the moduli space of the vector multiplets, and $\xi$ is
a characteristic scale related to the brane tension. 
It is easy to see that the potential vanishes
\begin{equation}
V_{SUGRA}=0
\end{equation}
leading to a static universe with
\begin{eqnarray}
\phi&=&-\frac{1}{\alpha}\ln (1-\alpha^2\xi\vert y\vert)\nonumber \\
a&=&(1-\alpha^2\xi\vert y\vert)^{1/4\alpha^2}\nonumber \\
\end{eqnarray}
where we have defined $dy=adx_5$. Notice that $b=1$ here.
This is a flat solution corresponding to a vanishing cosmological constant on the brane-world.

Let us explicitly show that $\Delta\Phi_2=0$ in supergravity. Indeed the scalar field is a solution
of the BPS equation springing from the requirement of bulk supersymmetry 
\begin{equation}
\frac{d\phi}{dy}\equiv \phi'=\nabla W.
\label{BPS}
\end{equation}
This equality is valid everywhere throughout the bulk.
This implies that $\Delta\Phi_2=0$ in supergravity.
 
\subsection{ Breaking Supersymmetry}

Since supersymmetry is not observed in nature, one should incorporate 
supersymmetry breaking. A natural way to break supersymmetry is  
by coupling the bulk scalar field to brane fields fixed at their vevs.
This leads to
\begin{equation}
U_B=TW
\end{equation}
where $T=1$ is the supersymmetric case. 
The new four dimensional potential becomes
\begin{equation}
V=\frac{(T^2-1)}{2}(W^2-(\nabla W)^2).
\end{equation}
The equations of motion are
\begin{equation}
H^2=-\frac{1}{16 a^4}\int d\tau \frac{da^4}{d\tau}(\dot \phi^2- (T^2-1)(W^2-(\nabla W)^2))
\label{h2}
\end{equation}
and the Klein-Gordon equation becomes
\begin{equation}
\ddot \phi + 4H\dot \phi =-\frac{T^2-1}{2}\nabla(W^2-(\nabla W)^2).
\end{equation}
Unlike the case discussed in the previous section, this equation is valid
throughout since the spatial derivatives of $\phi$ are related to derivatives
of the superpotential by supersymmetry in the bulk.
This follows from the fact that supersymmetry is only broken on the brane.
In the bulk the BPS condition 
(\ref{BPS}) is still valid. After a boost followed by a dilation the BPS condition becomes
\begin{equation}
\phi'=\nabla U_B
\end{equation}
throughout the bulk and on the brane\cite{us}. This implies that $\Delta\Phi_2=0$
when supersymmetry is broken on the brane.

Note that the breaking of supersymmetry on the brane leads to non-static solutions.
The explicit solution to these equations is obtained from the static solution
by going to conformal coordinates with $b=1$
\begin{equation}
ds^2=a^2(u)(-d\tau^2+du^2+dx^idx_i)
\end{equation}
and performing a boost along the $u$ axis
\begin{eqnarray}
a(u,\tau)&=& a(u+h\tau,\frac{\xi}{\sqrt {1-h^2}})\nonumber \\
\phi(u,\tau)&=& \phi (u+h\tau,\frac{\xi}{\sqrt {1-h^2}})\nonumber \\ 
\end{eqnarray}
where we have displayed the explicit $\xi$ dependence.
On the brane this implies that
\begin{eqnarray}
\dot a &=& ha' \nonumber \\
\dot \phi&=&h \phi'.\nonumber \\ 
\end{eqnarray}
Now for 
\begin{equation}
h=\pm \frac{\sqrt{T^2-1}}{T}
\end{equation} 
we find that 
\begin{equation}
\frac{da^4}{d\tau}(\dot \phi^2- (T^2-1)(W^2-(\nabla W)^2))=-h^2T^2\frac{d(a^4W^2)}{d\tau}
\end{equation}
which solves (\ref {h2}). Similarly the Klein-Gordon equation is satisfied.

The resulting universe is characterized by the scale factor
\begin{equation}
a(t)=T(1-\frac{\tau}{\tau_0})^{1/3+1/6\alpha^2}
\end{equation}
with $\tau_0=-(1/3+1/6\alpha^2)/(1/4-\alpha^2)h\xi$. Note that there is 
always a singularity, either in the past or in the future. 
As the metric is conformally flat one can derive the four dimensional Planck mass
obtained from dimensional reduction
\begin{equation}
m^2_{\rm p}=\frac{4M_5^3}{(2\alpha^2 +1)\xi}.
\end{equation}
The scale factor 
corresponds to a solution of the four dimensional FRW equations with an
acceleration parameter 
\begin{equation}
q_0=\frac{6\alpha^2}{1+2\alpha^2}-1
\end{equation}
and an equation of state
\begin{equation}
\omega_{SUGRA}=-1+\frac{4\alpha^2}{1+2\alpha^2}
\end{equation}
which never violates the dominant energy condition.
Notice that the universe is accelerating provided that 
\begin{equation}
-\frac{1}{2}\le\alpha \le \frac{1}{2}.
\end{equation}
The solution with $\alpha=-\frac{1}{\sqrt{ 12}}$ is accelerating, and the 
singularity recedes from the brane world \cite{us}. 
On the contrary the self-tuned model of \cite{dimopoulos} leads to a 
decelerating universe with the singularity converging towards the brane world.
It is also remarkable that a purely static solution exists too in that case.

The limit $\alpha\to 0$ is interesting as it corresponds to a free scalar 
field in the bulk together with a bulk cosmological constant. The coupling to 
the brane-world is constant too. The solution becomes
\begin{equation}
a(\tau)=T e^{h\xi\tau/4}.
\end{equation}
It is a pure exponential associated with a cosmological constant and 
$\omega_{SUGRA}=-1$.
The cosmological constant is identified with $3m_{\rm p}^2H^2$ leading to
\begin{equation}
\rho_{\Lambda}=\frac{3(T^2-1)}{T^2}\frac{\xi}{4\kappa_5^2}.
\end{equation}
It depends on the brane tension and the amount of supersymmetry breaking.
As in the four dimensional case, an acceptable value of the cosmological 
constant requires a fine-tuning of these parameters.

\section{Supergravity Cosmological Evolution}

Including matter on the brane results in the full Friedmann equation displayed
in (\ref{fried}) together with (\ref{h2}).  
In this section we examine the supergravity case in the linear regime where 
the $\rho^2$ is negligible in the Friedmann equation. 
\subsection{  Radiation Dominated Era}
We consider the supergravity case first with $V_{SUGRA}=0$.
In the radiation dominated era the Klein-Gordon equation reads
\begin{equation}
\ddot \phi + 4H\dot \phi=0
\end{equation}
combined with 
\begin{equation}
H^2=  \frac{1}{12}W\rho_m-\frac{1}{16 a^4}\int d\tau \frac{da^4}{d\tau}(\dot \phi^2).
\end{equation}
We find that
\begin{equation}
\phi=\phi_0
\end{equation}
is a solution leading to the usual expansion
\begin{equation}
a=a_0(\frac{t}{t_0})^{1/2}
\end{equation}
where $t_0$ is some initial time.
The Newton constant is identified with
\begin{equation}
\frac{8\pi G_N}{3}=\frac{\kappa_5^2 W(\phi_0)}{12}.
\end{equation}
The scalar field remains constant until the beginning of the matter dominated era.

\subsection{Matter Dominated Era}

In the matter dominated era the Klein-Gordon equation is modified
\begin{equation}
\ddot \phi + 4H\dot \phi +\frac{1}{6}\rho_m\nabla W=0.
\end{equation}
Similarly the Friedmann equation becomes
\begin{equation}
H^2= \frac{1}{12}W\rho_m-\frac{1}{16 a^4}\int d\tau \frac{da^4}{d\tau}\dot \phi^2
-\frac{1}{ 12 a^4}\int d\tau a^4\rho_m \frac{dW}{d\tau}.
\end{equation}
The solution to these equations is
\begin{eqnarray}
\phi&=&\phi_0+\beta\ln\frac{\tau}{\tau_e}\nonumber \\
a&=&a_e(\frac{\tau}{\tau_e})^{\gamma}\nonumber \\
\end{eqnarray}
where $\tau_e$ and $a_e$ are set at equality. 
We are interested in the small $\alpha$ case as it leads to an accelerating universe
when no matter is present and  small time deviations
for Newton's constant.

For small $\alpha$ we get 
\begin{eqnarray}
\beta&=&- \frac{8}{15}\alpha\nonumber \\
\gamma&=&\frac{2}{3}-\frac{8}{45}\alpha^2\nonumber\\
\end{eqnarray}
In a phenomenological way we identify Newton's constant with the ratio
\begin{equation}
\frac{8\pi G_N(\tau)}{3}\equiv \frac{H^2}{\rho_m}
\end{equation}
 We deduce that
\begin{equation}
\frac{G_N(\tau)}{G_N(\tau_e)}=(\frac{\tau}{\tau_e})^{-8\alpha^2/15}
\end{equation}
In terms of the red-shift $z$ this is 
\begin{equation}
\frac{G_N(z)}{G_N(z_e)}=(\frac{z+1}{z_e+1})^{4\alpha^2/5}
\end{equation}
For the supergravity case with $\alpha^2=1/12$ the exponent is $1/15$.
As $z_e\sim 10^3$ this leads to a decrease   by $37\%$ since equality.
For the self-tuned case of ref \cite{dimopoulos} $\alpha^2=1$ the exponent 
is $4/5$, leading to an even greater decrease since equality.

Notice that the Newton constant starts to decrease only from the time of
 matter and radiation equality and is
strictly constant during the radiation dominated era. Nucleosynthesis 
constrains the variation to be less than $20\%$. This implies that  
$\alpha\le 0.2$, see \cite{barrow} and references therein. This leads to the 
upper bounds  $q_0\le -0.77$ and $\omega_{SUGRA} \le -0.85$. Whilst our 
supergravity solution gave $\alpha^2$ to be $1/12$ one can imagine more 
complicated theories giving smaller values of $\alpha$. 

However, in our model we would expect the couplings to standard model
particles to also vary in a similar manner. This could lead to a variation
in, for example, the proton and neutron masses since these arise from Yukawa
couplings in the standard model. Since many of the 
tests for the variation of the Newton constant assume all other masses
and couplings are constant \cite{wetterich} it is possible that our 
supergravity variation
would evade detection. In order to ascertain this one would need to consider
the detailed cosmological perturbations predicted by the model. This is
currently in progress \cite{us&carsten}.

\subsection{Broken Supergravity}

If we denote by $H_{SUGRA}$ the Hubble parameter derived in the pure 
supergravity case, then 
the Friedmann equation in the broken supergravity case is
\begin{equation}
H^2=H^2_{SUGRA}+\frac{V}{8}
\end{equation}
where we have used the fact that $\phi$ varies slowly compared to $a$. 
The evolution coincides with the one obtained from unbroken  supergravity
as long as the contribution from the potential does not dominate.
In the radiation dominated era this requires
\begin{equation}
\frac{T^2-1}{T}\frac{W}{2\kappa^2_5}<<\frac{2}{3}\frac{\rho_e}{1-\alpha^2}
\end{equation}
where $\rho_e$ is the matter density at equality. This implies that the left-hand side
is much smaller that $10^{-39}$ GeV$^4$.
Let us now denote the supersymmetric brane tension by 
\begin{equation}
M_S^4=\frac{3W}{2\kappa_5^2}
\end{equation}
and the supersymmetry breaking contribution
\begin{equation}
M_{BS}^4=(T-1)M_S^4.
\end{equation}
We find that 
\begin{equation}
M^4_{BS}<<\frac{\rho_e}{1-\alpha^2}.
\end{equation}
Now this is an extreme fine-tuning of the non-supersymmetric contribution to the brane tension.

In the matter dominated era the supergravity Hubble parameter decreases faster than the 
potential contribution.
Coincidence between the matter dominated supergravity contribution
$H^2_{SUGRA}$ and the potential energy occurs at $\tau_c$ such that
\begin{equation}
(\frac{\tau_c}{\tau_e})^{3\gamma+\alpha\beta}=\frac{1}{1-\alpha^2}\frac{\rho_e}.
{M_{BS}^4}
\end{equation}
In terms of the red-shift this becomes
\begin{equation}
M^4_{BS}=\frac{1}{1-\alpha^2}(\frac{z_c+1}{z_e+1})^{3+\alpha\beta/\gamma}\rho_e.
\end{equation}
Imposing that coincidence has occurred only recently leads to 
\begin{equation}
M^4_{BS}\approx \rho_c
\end{equation}
where $\rho_c$ is the critical density.
It is relevant to reformulate this fine-tuning result in the $\alpha\to 0$ 
limit where
the universe expands due to a pure cosmological constant on the brane-world, we find 
\begin{equation}
\rho_{\Lambda}=M_{BS}^4.
\end{equation}
This is the usual extreme fine-tuning of the cosmological constant. Indeed it 
specifies that the energy density received by the brane-world from the 
non-supersymmetric sources, e.g.
radiative corrections and phase transitions, cannot exceed the critical energy density of the universe. 

Of course provided  this fine-tuning is  explained, which is not the case from
our five-dimensional perspective, the late evolution of the universe is the one described 
in section 5 with a cosmic acceleration characterized by $q_0$.

\section{Conclusions}

In this paper we have shown that a rich cosmology results from the scenario 
presented in \cite{us}. We have shown that the inclusion of matter on
the brane, bulk supergravity with supersymmetry breaking on the brane
results in the modified brane world FRW dynamics. Our brane world universe
undergoes the usual radiation dominated and matter dominated transitions.
In the matter dominated era, there is a variation in the Newton constant on 
the brane, though this is
probably accompanied by variation in the other fundamental constants
such that it evades detection. At late times the Friedmann equation
results in an accelerating universe for the bulk potential given by
supergravity. To obtain acceleration today we are required to fine-tune
the supersymmetry breaking parameter. This fine-tuning is not given by
the five dimensional physics, but is no more than is usually required in four dimensions.

We also considered a self-tuned brane, and obtained an oscillating  
universe at late times. Our solution here required a very small 
mass for the scalar field. This is to comply with the requirement
of a non-negligible dark energy component now. If this is not the case
the universe remains in a matter dominated phase.  The amplitude of the 
oscillations in Newton's constant and the couplings to standard model particles
and the Hubble parameter 
are sufficiently small that it is unlikely they 
could be detected. Nevertheless this model is ruled out as it would lead to 
rapid oscillations of the Hubble parameter now. 

In conclusion we have presented two typical scenarios in order to tackle the dark energy problem
from a five dimensional vantage. Both cases are reminiscent of quintessence models
where a scalar field  either rolls down a run-away potential or oscillates
around a minimum. Only the former leads to relevant four dimensional cosmology,
though we find that the fine-tuning of the cosmological constant has not been alleviated
by going to five-dimensions. It seems difficult to foresee a likely alternative to this result,
although some results along these lines have been presented recently\cite{gro}.

Barring the fine-tuning of the supersymmetry breaking we expect that the 
model \cite{us}
coupled to matter deserve to be further studied. In particular the 
analysis of perturbations should be  fruitful \cite{us&carsten}.

\acknowledgments
We wish to thank J. Barrow, C. van de Bruck, J.P. Uzan and C. Wetterich for 
discussions. PhB thanks DAMTP and ACD thanks CERN for hospitality whilst this 
work was in progress. This work was supported in part by PPARC.

\def\Journal#1#2#3#4{{#1}{\bf #2}, #3 (#4)}
\def\NPB{Nucl.\ Phys.\ {\bf B}}
\def\PLB{Phys.\ Lett.\ {\bf B}}
\def\PRL{Phys.\ Rev.\ Lett.\ }
\def\PRD{Phys.\ Rev.\ D }
\def\JPA{J.\ Phys.\ {\bf A}}
\def\JHEP{JHEP}

\end{document}